\documentclass[prd,aps,twocolumn,preprintnumbers,superscriptaddress,showpacs,floatfix]{revtex4-1}
\pdfoutput=1
\usepackage{amsmath}
\usepackage{amsfonts}
\usepackage{mathtools}
\usepackage{multirow}
\usepackage{microtype}
\usepackage{hyperref}
\hypersetup{
 colorlinks,
 linkcolor={red!50!black},
 citecolor={blue!50!black},
 urlcolor={blue!80!black}
}
\usepackage{tikz}
\usepackage{booktabs}
\usepackage[autostyle=true]{csquotes}
\usepackage[toc,page]{appendix}

\newtheorem{example}{Example}
\newcommand{\myparagraph}[1]{\bigskip\vspace{-0.5em}\noindent\textbf{#1}}
\newcommand{\CC}{\mathbb{C}}
\newcommand{\ZZ}{\mathbb{Z}}
\newcommand{\QQ}{\mathbb{Q}}
\newcommand{\PP}{\mathbb{P}}

\begin{document}
\title{Hilbert Series, Machine Learning, and Applications to Physics}
\author{Jiakang Bao}
\email{jiakang.bao@city.ac.uk}
\affiliation{Department of Mathematics, City, University of London, EC1V 0HB, UK}
\author{Yang-Hui He}
\email{hey@maths.ox.ac.uk}
\affiliation{Department of Mathematics, City, University of London, EC1V 0HB, UK}
\affiliation{Merton College, University of Oxford, OX1 4JD, UK}
\affiliation{School of Physics, NanKai University, Tianjin, 300071, P.R. China}
\author{Edward Hirst}
\email{edward.hirst@city.ac.uk}
\affiliation{Department of Mathematics, City, University of London, EC1V 0HB, UK}
\author{Johannes Hofscheier}
\email{johannes.hofscheier@nottingham.ac.uk}
\affiliation{School of Mathematical Sciences, University of Nottingham, Nottingham, NG7 2RD, UK}
\author{Alexander Kasprzyk}
\email{a.m.kasprzyk@nottingham.ac.uk}
\affiliation{School of Mathematical Sciences, University of Nottingham, Nottingham, NG7 2RD, UK}
\author{Suvajit Majumder}
\email{suvajit.majumder@city.ac.uk}
\affiliation{Department of Mathematics, City, University of London, EC1V 0HB, UK}
\begin{abstract}
We describe how simple machine learning methods successfully predict geometric properties from Hilbert series~(HS). Regressors predict embedding weights in projective space to~${\sim}1$ mean absolute error, whilst classifiers predict dimension and Gorenstein index to~$>90\%$ accuracy with~${\sim}0.5\%$ standard error. Binary random forest classifiers managed to distinguish whether the underlying~HS describes a complete intersection with high accuracies exceeding~$95\%$. Neural networks~(NNs) exhibited success identifying~HS from a Gorenstein ring to the same order of accuracy, whilst generation of ``fake''~HS proved trivial for~NNs to distinguish from those associated to the three-dimensional Fano varieties considered.
\end{abstract}
\maketitle
\section{Introduction and Summary}\label{intro}
The Hilbert series~(HS) is an important invariant in the study of modern geometry. In physics,~HS have recently become a powerful tool in high energy theory, appearing, for example, in the study of: Bogomol'nyi–Prasad–Sommerfield~(BPS) operators of supersymmetric gauge theories~\cite{Benvenuti:2006qr,Feng:2007ur}; supersymmetric quantum chromodynamics~(SQCDs)~\cite{Gray:2008yu,Hanany:2008sb,Chen:2011wn,Jokela:2011vg} and instanton moduli spaces~\cite{Benvenuti:2010pq,Hanany:2012dm,Buchbinder:2019eal}; invariants of the standard model~\cite{Hanany:2010vu,Lehman:2015coa}; polytopes which arise in string compactifications~\cite{Braun:2012qc}; and explicit constructions of effective Lagrangians~\cite{Lehman:2015via,Henning:2015daa,Kobach:2017xkw,Anisha:2019nzx,Marinissen:2020jmb,Graf:2020yxt}.

In parallel, a programme to use machine learning~(ML) techniques to study mathematical structures has recently been proposed~\cite{He:2017aed,He:2018jtw,DavEtAl21}. The initial studies were inspired by timely and independent works~\cite{He:2017aed,He:2017set,Krefl:2017yox,Ruehle:2017mzq,Carifio:2017bov}. In these, the effectiveness of~ML regressor and classifier techniques in various branches of mathematics and mathematical physics has been investigated. Applications of~ML include: finding bundle cohomology on varieties~\cite{Ruehle:2017mzq,Brodie:2019dfx,Larfors:2020ugo}; distinguishing elliptic fibrations~\cite{He:2019vsj} and invariants of Calabi--Yau threefolds~\cite{Bull:2018uow}; the Donaldson algorithm for numerical Calabi--Yau metrics~\cite{Ashmore:2019wzb}; the algebraic structures of groups and rings~\cite{He:2019nzx}; arithmetic geometry and number theory~\cite{Alessandretti:2019jbs,He:2020eva,He:2020kzg}; quiver gauge theories and cluster algebras~\cite{Bao:2020nbi}; patterns in particle masses~\cite{Gal:2020dyc}; statistical predictions and model-building in string theory~\cite{Deen:2020dlf,Halverson:2019tkf,Halverson:2020opj}; and classifying combinatorial properties of finite graphs~\cite{He:2020fdg}. Here we apply~ML techniques to the plethystic programme of using Hilbert series to understand structures of quantum field theory.
The physical motivation for this work has two primary applications. First, when considering a generic supersymmetric quantum field theory the number of BPS operators at each order is given by the initial terms in the Hilbert series. Computing these operator frequencies requires significant computational power, particularly for higher order terms (for the multi-trace case the growth is exponential). In this work the goal for the machine learning techniques implemented is to return information about the full series' closed form, which can then directly provide the higher order information, hence bypassing the need for order-by-order computation.
Second, from a string perspective the geometry of the moduli space has an array of physical applications and if these techniques can return the underlying variety's geometric properties directly the vacuum can be analysed without need for complete information about the theory.

We examined databases of~HS arising in geometry -- see~\cite{ABR:Fano,BK22} and the Graded Ring Database~(GRDB)~\cite{grdb} -- and ``fake''~HS generated to imitate the ``real'' geometric~HS. Simple~ML methods were able to successfully predict several geometric quantities associated to the~HS, and were able to accurately distinguish real from fake~HS.

Depending on the form of the~HS, simple regression neural networks~(NNs) managed to learn the embedding weights in projective space to mean absolute error~(MAE)~${\sim}1$; whilst classification~NNs predicted the dimension and Gorenstein index with both accuracy and Matthews correlation coefficient~(MCC) in excess of~$0.9$.

Motivated by the question of whether~ML can detect when a~HS comes from a Gorenstein ring, we found that binary classifiers identified whether a fake~HS had a palindromic numerator to accuracy and~MCC greater than~$0.9$. Binary classifiers were easily able to distinguish the fake generated data from the dataset of~HS associated to three-dimensional Fano varieties obtained from~\cite{grdb,fanodata}.

A random forest classifier correctly predicted whether the~HS described a complete intersection~(CI): this was achieved with accuracy~$0.9$ and~MCC~$0.8$ when the numerator (padded with~$0$'s) of the~HS was used as input; and with accuracy~$0.95$ and~MCC greater than~$0.9$ when the Taylor series (to order~$100$) of the~HS was used.

Code scripts for these investigations, along with the datasets generated and analysed, are available from:
\begin{center}
\url{https://github.com/edhirst/HilbertSeriesML.git}
\end{center}

\section{Hilbert Series and Physics}\label{hs}
The~HS is an important quantity that encodes numerical properties of a projective algebraic variety. It is not a topological invariant in that it depends on the embedding under consideration~\cite[Example 13.4]{Harris}. We work throughout with varieties defined over~$\CC$.

Given a complex projective variety~$X$ and ample divisor~$D$ there exists a natural embedding in a weighted projective space (w.p.s.)~$\PP_\CC(p_0,\ldots,p_k)$. We denote its homogeneous coordinate ring by~$R$, i.e.~$R=\CC[X_0,\ldots,X_k]/I$ where the variables~$X_i$ have weights~$p_i$, and~$I$ is the homogeneous ideal generated by the polynomials defining~$X$. We write~$\PP_\CC(p_0^{q_0}, \ldots, p_s^{q_s})$ as shorthand to indicate that the weight~$p_i$ appears~$q_i$ times. The embedding of $X$ into the w.p.s.\ induces a grading on~$R=\bigoplus_{i\ge0} R_i$. We refer to~\cite{Dolga} for details.

The~HS is the generating function for the dimensions of the graded pieces of~$R$:
\[
H(t; X) = \sum\limits_{i=0}^{\infty} (\dim_{\CC} R_i) t^i
\]
where~$\dim_{\CC}R_i$, the dimension of the~$i$-th graded piece of the ring~$R$, can be thought of as the number of independent degree~$i$ polynomials on the variety~$X$. The map~$i\mapsto\dim_{\CC} R_i$ is called the~\emph{Hilbert function}.

By the Hilbert--Serre Theorem (see for example~\cite[Theorem 11.1]{AtiyahMacdonald}) there exists~$P \in \ZZ[t]$ such that
\begin{equation}\label{refined_HS}
H(t; X) = \frac{P(t)}{\prod\limits_{i=0}^s(1-t^{p_i})^{q_i}}. 
\end{equation}
Let~$j$ be the smallest positive multiple such that~$jD$ is very ample. We call~$j$ the~\emph{Gorenstein index}, and can rewrite~\eqref{refined_HS} in the form:
\begin{equation}\label{refined_HS_Palin}
H(t; X) = \frac{\tilde{P}(t)}{(1-t^j)^{\dim+1}}
\end{equation}
Here~$\dim$ is the dimension of~$X$, and~$\tilde{P} \in \ZZ[t]$. If~$R$ is a Gorenstein graded ring then the numerator is a palindromic polynomial (by Serre duality). Recall that a polynomial~$\sum_{i=1}^d a_i t^i$ is called palindromic if~$a_i=a_{d-i}$~\cite{Stanley1978}.

For example, consider the complex line~$M=\CC$ (regarded as the affine cone over a point) parameterised by a single complex variable~$x$. Then the~$i$-th graded piece~$R_i$ is generated by the single monomial~$x^i$. Thus,~$\dim_{\CC} R_i = 1$ for all~$i \in \ZZ_{\geq 0}$ so that the~HS becomes~$H(t; \CC) = (1-t)^{-1}$. In general, we have that~$H(t; \CC^n) = (1-t)^{-n}$.

\myparagraph{The Plethystic Programme.}
In supersymmetric gauge theories, when the vevs of scalars in different supermultiplets are turned on, the (vacuum) moduli spaces are non-trivial algebraic varieties~\cite{vms,LT,Mehta:2012wk} such as hyperk\"ahler cones and (closures of) symplectic leaves. In this case~HS are a powerful tool to enumerate gauge invariant operators~(GIOs) at different orders.

A particularly useful application of~HS to theoretical physics is the plethystic programme, which reveals more information of the moduli spaces. We leave a detailed summary of the key formulae to Appendix~\ref{ap:pleth}.

The multi-graded~HS, i.e.\ the multi-variate series
\[
H(t_1,\dots,t_k; X) = \sum\limits_{\vec{i}=0}^{\infty} \dim_{\mathbb{C}}(X_{\vec{i}} ) t_1^{i_1}\dots t_k^{i_k}
\]
obtained by considering multi-graded rings with pieces~$X_{\vec{i}}$ for~$\vec{i}=(i_1,\dots,i_k)$, could fully determine how the GIOs transform under symmetry groups of gauge theories.

\myparagraph{Duality and Moduli Spaces.}
HS have been well-studied in the context of quiver gauge theories. For Higgs branches in low dimensions,~HS obtained from the Molien--Weyl integral enable us to systematically study the geometry of~SQCDs~\cite{Gray:2008yu}. Such methods can also be used to study the instanton moduli spaces~\cite{Benvenuti:2010pq,Hanany:2012dm,Dey:2013fea}. As the spaces of dressed monopole operators, i.e.\ the Coulomb branches, receive quantum corrections, monopole formula~\cite{Cremonesi:2013lqa} and Hall--Littlewood formula~\cite{Cremonesi:2014kwa} are used to obtain the~HS. This not only unveils the geometry of moduli spaces, but also provides tools and evidences to study three-dimensional mirror symmetry and duality including theories in higher dimensions.

\myparagraph{Standard Model.}
Phenomenologically,~HS have been applied to lepton and quark flavour invariants for the Standard Model in~\cite{Hanany:2010vu} as well as to the minimal supersymmetric Standard Model in~\cite{He:2014loa,Xiao:2019uhh}.

\section{Machine Learning}\label{ml}
In this section we describe our approaches to~ML properties of the rational representations~\eqref{refined_HS} and~\eqref{refined_HS_Palin} by feeding in coefficients of the corresponding~HS. \texttt{Keras} with the~\texttt{TensorFlow} backend~\cite{Tensorflow} was used for the investigations. In~\S\ref{Real_HS_section}, ``real''~HS associated to certain three-dimensional Fano varieties are introduced and analysed. In~\S\ref{Fake_HS_section}, ``fake''~HS, i.e.\ rational functions of the form~\eqref{refined_HS} and~\eqref{refined_HS_Palin}, were generated and properties of them were machine-learnt. In~\S\ref{detectpalin} and~\S\ref{detectreal}, binary classifiers were used to determine whether fake~HS of the form~\eqref{refined_HS_Palin} had palindromic numerator, and to determine fake~HS from real~HS, all with great success. Finally, in~\S\ref{ci} we use~ML to determine if a~HS is associated to a complete intersection.

\subsection{Acquiring~HS}\label{Real_HS_section}
Example~HS associated to algebraic varieties were retrieved from the~GRDB~\cite{grdb,fanodata}. We use a database of candidate~HS conjecturally associated to three-dimensional~$\QQ$-Fano varieties with Fano index one, as constructed in~\cite{ABR:Fano,BK22}. Such varieties come with a natural choice of ample divisor~$D=-K$, the anti-canonical divisor. We call these~HS ``real''.
See Appendix~\ref{real_HS_distributions} for the distributions of the parameters~$d,\{a_i\},s,\{p_\ell\},\{q_\ell\}$ for this set of data. Here we are using notation as in~\eqref{refined_HS}, and write~$P(t) = 1+\sum_{i=1}^d a_i t^i$ for the numerator polynomial.

\begin{example}\label{ex:Gorenstein}
Consider the three-dimensional~$\QQ$-Fano variety~$X\subset\PP(1^3,2^2,3^2)$ (number~$11122$ in the~GRDB). This is of codimension~$3$, with~$\mathcal{B}=\{\frac{1}{2}(1,1,1),2\times\frac{1}{3}(1,1,2)\}$ isolated orbifold points, and hence has Gorenstein index~$j=6$. Writing the~HS in the form~\eqref{refined_HS} gives:
\begin{align*}
H(t;X)=\frac{P(t)}{(1-t)^3(1-t^2)^2(1-t^3)^2}&\\
\text{ where }P(t)=1-2t^4-2t^5+&2t^7+2t^8-t^{12}.
\end{align*}
Rewriting this in the form~\eqref{refined_HS_Palin} gives:
\begin{align*}
H(t;X)=\frac{\tilde{P}(t)}{(1-t^6)^4}&\\
\text{ where }\tilde{P}(t)=1+&3t+8t^2+\ldots+8t^{21}+3t^{22}+t^{23}.
\end{align*}
\end{example}

For the~HS of this dataset, there are two competing phenomena that contribute to its coefficients: the initial part~$P_{\mathrm{ini}}$ that coincides with the~HS in small degrees and the ``correction terms''~$P_{\mathrm{orb}}(Q)$ for each isolated orbifold point~$Q=\frac1r(b_1,\ldots,b_{\dim})$ of~$X$.
More precisely, we have~\cite{icecream}
\[
 H(t; X) = P_{\mathrm{ini}} + \sum_{Q\in\mathcal{B}} P_{\mathrm{orb}}(Q)
\]
where the sum is taken over the set~$\mathcal{B}$ of isolated orbifold points of~$X$. $P_{\mathrm{ini}}$ and~$P_{\mathrm{orb}}(Q)$ ($Q=\frac1r(b_1,\ldots,b_{\dim})$) satisfy
\[
P_{\mathrm{ini}} = \frac{A(t)}{(1-t)^{\dim+1}},\quad P_{\mathrm{orb}}(Q)=\frac{B_Q(t)}{(1-t)^{\dim}(1-t^r)}
\]
where~$A(t), B_Q(t)$ are integral palindromic polynomials with degrees related via~$\deg{B_Q(t)}-\deg{A(t)} = r-1$. The coefficients (called \emph{plurigenera}) of the~HS of~$H$ coincide with~$P_{\mathrm{ini}}$ in degrees~$\le\lfloor\deg{A(t)}/2\rfloor$, whilst in higher degrees the orbifold points start to contribute to the plurigenera. Because of this phenomenon, extra care must be taken when computing parameters for the representations~\eqref{refined_HS} and~\eqref{refined_HS_Palin} from a finite set of coefficients of the~HS. Our investigations show that~ML can cope with this behaviour.

\subsection{Generating and~ML Fake~HS}\label{Fake_HS_section}
The ``fake''~HS generated take the forms~\eqref{refined_HS} and~\eqref{refined_HS_Palin}, with numerators of the form~$1+\sum_{i=1}^d a_i t^i$. The numerator~$\tilde{P}(t)$ of~\eqref{refined_HS_Palin} is required to be palindromic (and, as a consequence,~$a_d = 1$). Coefficient sets consisting of the parameters~$d, \{a_i\}, s, \{p_\ell\}, \{q_\ell\}$, where~$1\leq i\leq d$ and~$1\leq\ell\leq s$, were randomly generated and the Taylor expansions of the resulting fake~HS were computed to order~${\sim}1000$. If the parameters did not satisfy~$\sum_\ell p_\ell q_\ell> d$, if there were negative coefficients in the resulting Taylor expansion, or if they matched a real Hilbert series then the parameters were discarded.

The resulting data were fed into a~NN to learn the desired properties of the fake~HS. The input was a vector of Taylor expansion coefficients: either a vector of coefficients for low-order terms~$0$ to~$100$; or for high-orders terms~$1000$ to~$1009$. Although coefficients of low-order terms are easier to calculate, predictions based on those inputs are more error-prone as contributions from orbifold points take effect only for high-order terms (see~\S\ref{Real_HS_section}).

Fewer coefficients were required when learning from coefficients deeper in the Taylor expansion; geometric data are more readily extracted from larger plurigenera. We found the following analogy from toric geometry insightful. When counting the number of lattice points~$c_m=\left|m\Delta\cap\ZZ^{\dim}\right|$ in the~$m$-th dilation of a polytope~$\Delta$ then, for~$m\gg0$,~$c_m\sim\mathrm{Vol}(m\Delta)=m^{\dim}\mathrm{Vol}(\Delta)$. (This is a toric rephrasing of the~HS, with~$\Delta$ the polytope associated with an ample divisor~$D$ and~$c_m=h^0(mD)$.)

The first investigation used supervised regressor~NNs to learn~$\{p_\ell\}$ for fake~HS in the form~\eqref{refined_HS}. Supervised classifier~NNs were trained to predict the Gorenstein index~$j$ and the dimension~$\dim$ of fake~HS in the form~\eqref{refined_HS_Palin}. Classifiers were used since the~NN outputs were single numbers and hence associated well to classifier data structures.

\begin{figure}[!h]
\centering
\includegraphics[width=7cm]{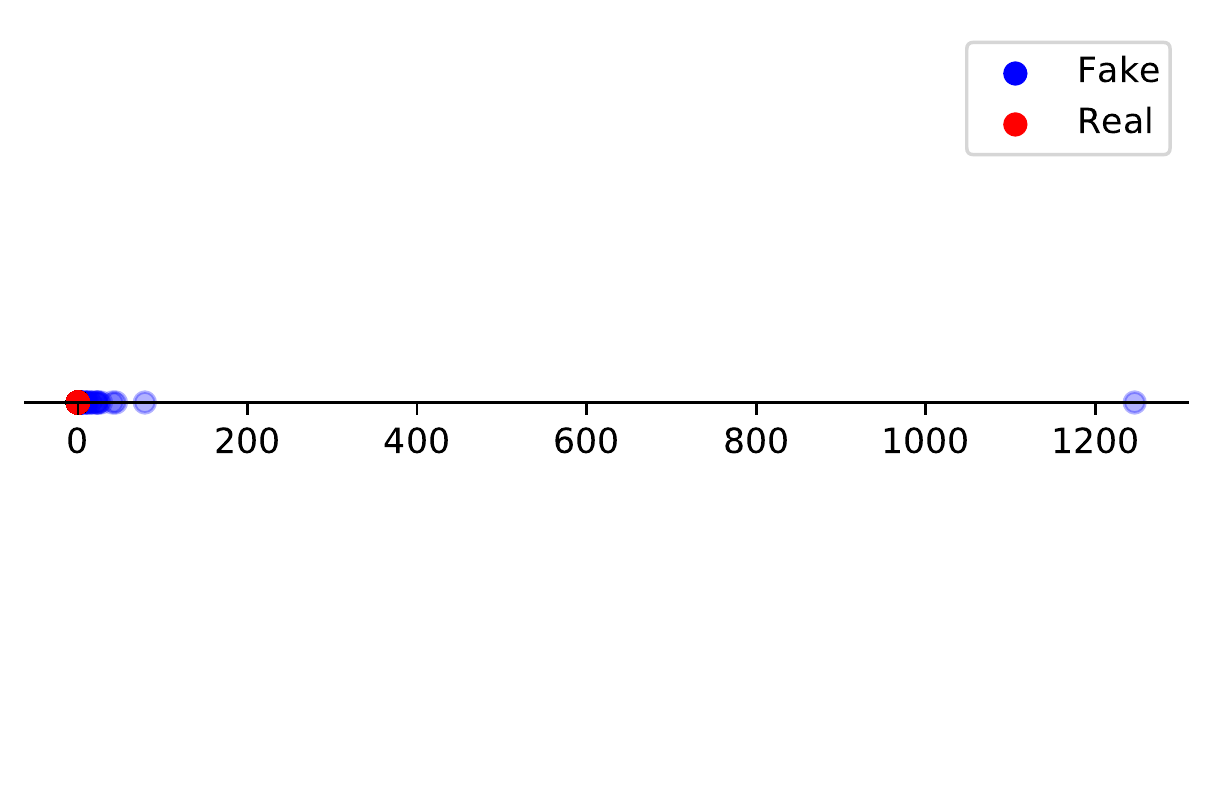}
\caption{PCA for~HS Taylor expansion coefficients coming from the~GRDB, \textit{`Real'}, or those randomly generated, \textit{`Fake'}.\label{realfake_pca}}
\end{figure}

We conclude with a comparison of the collected fake~HS data with the real~HS data from the~GRDB. We use the unsupervised method of principal component analysis~(PCA) to project the classes onto the highest variance linear component (see Figure~\ref{realfake_pca}). The PCA was performed on the vectors of the first 100 coefficients, with prior scalar transformation. The explained variance ratios give the normalised eigenvalues for the covariance matrix, sorted into a decreasing order. For the fake to real HS comparison the first eigenvalue ($0.78$) was significantly larger than the second ($0.16$) and subsequent 98 eigenvalues ($<0.04$). This indicates that one principal component is sufficient for description of the data distribution, and this principal component pays linearly progressively more attention to coefficients throughout the input HS vector up to the 24th where it then considers equal contributions from the remaining coefficients.

The projection shows a separation between the classes, indicating that there is linear structure in the data. Despite great efforts we were unable to break this separation. This raises the following question, to which we do not currently know the answer: what additional properties do fake~HS need to satisfy to better approximate the~GRDB~HS data?

\myparagraph{HS Regressor Investigations.}
For this investigation~${\sim}10\,000$ fake~HS of the form~\eqref{refined_HS} were uniformly drawn from a sample space given by~$d=3$,~$s=3$,~$|a_i|\le 10$,~$p_\ell\le 10$. This space was chosen to provide a sufficiently large range of fake~HS whilst ensuring that its size was still feasible for ML~training. The goal was to predict the values~$\{p_\ell\}$ and~$\{q_\ell\}$ of the form~\eqref{refined_HS} from a given (finite) range of~HS coefficients. This information was encoded into a single vector where each~$p_\ell$ was repeated~$q_\ell$ times, and the entries were given in increasing order.

A $5$-fold cross-validation (in the sense of~\cite{friedman2001elements}) was performed for a feed-forward regressor~NN with~$4$ hidden dense layers of~$1024$ neurons each, using LeakyReLU activation (with~$\alpha = 0.01$), in batches of~$32$ for~$20$ epochs over the full dataset. 
The~NN had a final dense layer with as many neurons as~$p_\ell$'s (counting multiplicities).
Dropout layers between the dense layers reduced the risk of overfitting (dropout factor~$0.05$).
The~NN was trained with the Adam optimiser~\cite{Adam} using a~\emph{log(cosh)} loss function and the training performance was measured via~MAE.
\emph{log(cosh)} is a continuous version of MAE used as the loss function such that training performance would be improved for gradient descent near the MAE discontinuity, however MAE provides a more interpretable metric of learning performance so is used as the metric on the independent test data.

Table~\ref{RefinedHSRegressorResults} summarises the averaged~MAE, with standard error, over the~$5$-fold cross validation for two ranges of~HS coefficients: the first~$101$ coefficients; and the coefficients of order~$1000$ to~$1009$. In both cases the~MAE is below~$2$, i.e.\ the true denominator of the form~\eqref{refined_HS} of the underlying~HS could be extracted with reasonably good accuracy from the~HS coefficients alone.

\begin{table}[tb]
\centering
\setlength{\tabcolsep}{0.5em}
\begin{tabular}{cc}
\toprule
Orders of Input & MAE\\
\cmidrule{1-2}
$0$ to~$100$&$1.94 \pm 0.11$\\
$1000$ to~$1009$&$1.04 \pm 0.12$\\
\bottomrule
\end{tabular}
\caption{Averaged MAE, with standard error, of the $5$-fold cross-validation of the~NN learning the weights~$p_\ell$ (with multiplicity) of the form~\eqref{refined_HS} of the~HS from input vectors of~HS coefficients to the specified orders.}
 \label{RefinedHSRegressorResults}
\end{table}

\myparagraph{HS Classification Investigations.}
In this investigation a $5$-fold cross-validation for a feed-forward classifier~NN with the same layer structure as before was trained. We again used an Adam optimiser, but now with~\emph{sparse categorical cross entropy} loss to reflect the classification question. Training performance was measured with accuracy and~MCC. The final dense layer now had as many neurons as classes in the investigation (5 in both cases), with softmax activation, and neurons representing the values the learnt parameters could take.

This time~${\sim}10\,000$~HS of the form~\eqref{refined_HS_Palin} were uniformly drawn from a sample space given by~$d = 5$,~$|a_i| \le 50$,~$j \le 5$,~$\dim \le 5$. The goal this time was to train an~NN to predict the Gorenstein index~$j$, the dimension~$\dim$, and the form~\eqref{refined_HS_Palin} from the~HS coefficients in the same orders of degrees.

Note if coefficients in larger degrees were used as input, the larger values caused problems with the loss function. This issue was mitigated by log-normalising the~HS coefficients, i.e.\ by taking the natural logarithm input values were scaled down to ranges the loss function and optimiser could handle. However some fake~HS contained~$0$ coefficients and were therefore omitted, hence resulting in a full dataset of $8711$~HS for the training with~HS coefficients of larger degree. Note also that log-normalisation was only used in this case and in no other investigations.

Table~\ref{PalinRefinedHSClassifierResults} summarises the averaged accuracies and~MCCs, with standard error, over the $5$-fold cross-validation of the~NN. These results show almost perfect classification of both the Gorenstein index,~$j$, and the dimension, dim, from~HS coefficients in low degrees. Interestingly the performance is worse when using terms deeper in the~HS, presumably due to the required log-normalisation of the coefficients removing the finer structure of the coefficients required to determine the exact parameter value being learnt.

\begin{table}[tb]
\centering
\setlength{\tabcolsep}{0.5em}
\begin{tabular}{cccc}
\toprule
Parameter & Orders & \multicolumn{2}{c}{Performance Measures}\\
Learnt & of Input & Accuracy & MCC\\
\cmidrule{1-4}
\multirow{2}{*}{$j$} & $0$ to~$100$ & $0.934 \pm 0.008$ & $0.916 \pm 0.010$ \\
& $1000$ to~$1009$ & $0.780 \pm 0.018$ & $0.727 \pm 0.022$ \\
\cmidrule{1-4}
\multirow{2}{*}{dim} & $0$ to~$100$ & $0.995 \pm 0.005$ & $0.993 \pm 0.006$ \\
 & $1000$ to~$1009$ & $0.865 \pm 0.024$ & $0.822 \pm 0.031$\\
\bottomrule
\end{tabular}
\caption{Averaged accuracy and~MCC, with standard error, of the $5$-fold cross-validation of the~NN learning the Gorenstein index~$j$, the dimension~$\dim$, and the~form~\eqref{refined_HS_Palin} with~HS coefficients in the specified ranges of degrees as input.}
\label{PalinRefinedHSClassifierResults}
\end{table}

\subsection{Identifying the Gorenstein Property}\label{detectpalin}
In this section we investigate the effectiveness of binary classifiers to detect if the numerator of form~\eqref{refined_HS_Palin} of an~HS is palindromic. Recall from Section~\ref{hs} that the numerator is palindromic if the ring~$R$ is Gorenstein (by Serre duality). Then the numerator of form~\eqref{refined_HS} is palindromic too (possibly up to a sign); see Example~\ref{ex:Gorenstein} for an illustration. The goal was to use a~NN to distinguish whether a~HS is coming from a Gorenstein ring, i.e.\ the numerator polynomial of form~\eqref{refined_HS_Palin} is palindromic. As before the~NN's input were~HS coefficients from the same ranges of degrees.

For the investigation two equally sized sets of fake~HS, one with and the other without palindromic numerators, were uniformly drawn from a sample space given by~$d = 9$,~$|a_i| \leq 50$,~$j = 5$,~$\dim+1 = 6$. The same reasons as before apply for this choice of space. The~HS in each of the two sets were then labelled and together comprised the full dataset for a $5$-fold cross-validation to be performed using a feed-forward classifier~NN with the same layer structure as in the previous investigation. Also the same Adam optimiser was used for training, but now with~\emph{binary cross-entropy} loss to reflect the classification question. Training performance was measured with accuracy and~MCC. The final dense layer of the~NN now had~$2$ neurons corresponding to whether the~HS comes from a Gorenstein ring or not.

Table~\ref{PalinBinaryClassifierResults} summarises the averaged accuracies and~MCCs, with standard error, over the $5$-fold cross-validation of the~NN. The results show good success in detecting if a~HS comes from a Gorenstein ring using~HS coefficients alone. The classifier performed better on coefficients in larger degrees indicating that the palindromicity property is more readily evident from plurigenera deeper in the~HS (possibly because of the bigger variation).

\begin{figure}[tb]
\centering
\includegraphics[width=7cm]{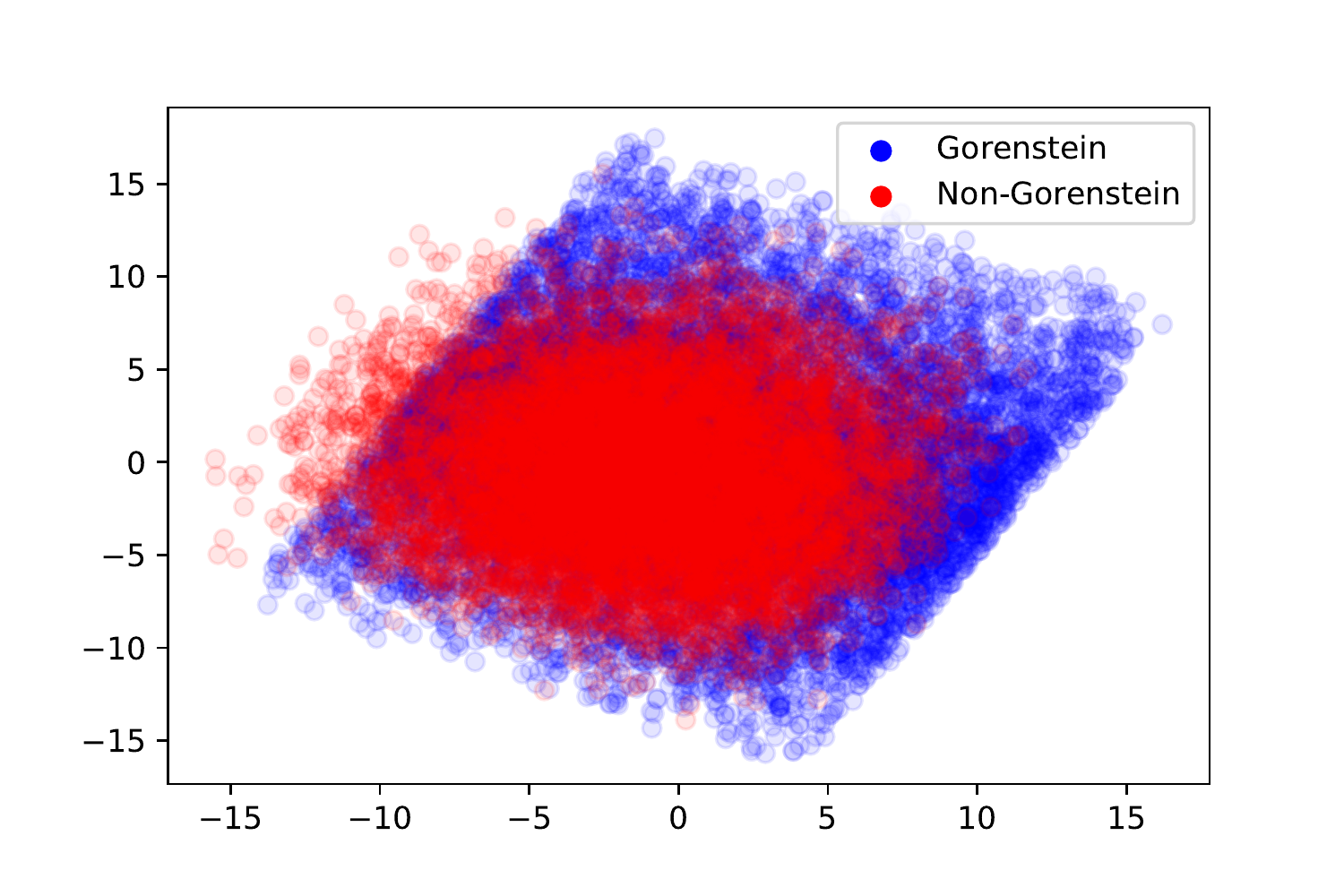}
\caption{The PCA for~HS Taylor expansion coefficients corresponding to~HS defined over Gorenstein rings or non-Gorenstein rings.}\label{gorenstein_pca}
\end{figure}

In addition,~PCA was also applied to the data in this binary classification problem, as seen in Figure~\ref{gorenstein_pca}, with similar behaviour for both low and high orders of input. This figure (for the low order inputs) highlights a lack of linear structure which the architecture could take advantage of. 
The PCA explained variance ratios for the 101 low order inputs show equal importance of the first two principal components (0.29, 0.27), lower importance for the next three components (0.19, 0.11, 0.10), minimal importance of the next four components ($\sim 0.01$), then negligible contribution from the remaining 92 ($\lesssim 10^{-30}$).
Equivalently for the high order inputs the first two components are dominant (0.30,0.26), with the next three less important (0.20,0.12,0.10), and the remaining five negligible ($\lesssim 10^{-10}$).
In both cases the two dominant principal components have a mix of contributions from components with no discernible pattern across the HS vector of coefficients. The full outputs can be observed in this paper's respective GitHub scripts.

\begin{table}[tb]
\centering
\setlength{\tabcolsep}{0.5em}
\begin{tabular}{ccc}
\toprule
Orders & \multicolumn{2}{c}{Performance Measures}\\
of Input & Accuracy & MCC\\
\cmidrule{1-3}
$0$ to~$100$ & $0.844 \pm 0.087$ & $0.717 \pm 0.155$\\
$1000$ to~$1009$ & $0.954 \pm 0.043$ & $0.919 \pm 0.073$\\
\bottomrule
\end{tabular}
\caption{Averaged accuracy and~MCC, with standard error, of the $5$-fold cross-validation of a~NN learning whether the~HS has palindromic numerator in form~\eqref{refined_HS_Palin} from~HS coefficients to the specified orders as input.}
\label{PalinBinaryClassifierResults}
\end{table}

\subsection{Differentiating Real and Fake~HS}\label{detectreal}
This investigation examined the success of a binary classifier in distinguishing whether a~HS, represented by a finite set of~HS coefficients, corresponds to a real~HS from the~GRDB, or a randomly generated fake~HS. The dataset consisted of~HS candidates conjecturally associated to~$3$-dimensional Fano polytopes from the~GRDB, amounting to~${\sim}29\,000$~HS, along with as many fake~HS with the same structure which were randomly generated.

A~$5$-fold cross-validation for a feed-forward classifier~NN with the same layer structure as in the previous investigations was performed. For training an Adam optimiser with a binary cross-entropy loss with the same parameters as before was used. Training performance was measured with accuracy and~MCC. The final dense layer had~$2$ neurons corresponding to whether the inputted~HS coefficients were associated to a real or fake~HS.

The~${\sim}29\,000$ fake~HS were generated randomly using form~\eqref{refined_HS} parameters drawn from probability distributions reflecting the real~HS data as given in Appendix~\ref{real_HS_distributions}. An equal number of real~HS were taken from the~GRDB to produce the full dataset, and as before~HS coefficients to the same order of degrees were used as~NN inputs.

In this investigation the averaged accuracies and~MCCs exceeded~$0.99$ for both ranges of degrees of~HS coefficients. Further analysis of the data showed that coefficients of fake~HS were orders of magnitudes different to the real case which possibly made this classification far easier. Resampling such that the coefficients were more comparable, although improving this investigations complexity, would make the fake data less representative with respect to the underlying variety's properties. Hence we chose to use the same data throughout all investigations despite this binary classification becoming more trivial; as corroborated by the 1d PCA separation in Figure \ref{realfake_pca}. This also highlights the uniqueness of real~HS which come with a wealth of further impactful structure, e.g.\ on the parameters of the corresponding forms~\eqref{refined_HS} and~\eqref{refined_HS_Palin}.

\subsection{Detecting Complete Intersection}\label{ci}
An important application of the plethystic logarithm (see Appendix~\ref{ap:pleth} for details and references) is that it detects whether the underlying variety is a~\emph{complete intersection}~(CI), i.e.\ the defining ideal (the ideal of polynomials vanishing on the variety) is generated by exactly codimension many polynomials. Such optimal intersection has been widely used in the physics literature, e.g.\ in string model-building~\cite{Candelas:1987kf,Anderson:2007nc}. As can be seen from the definition, the~$\text{PE}^{-1}$ involves the number-theoretic~$\mu$-function, making the computation non-trivial. A natural question arises as to whether a trained classifier can identify whether~$X$ is CI, i.e.\ when~$\text{PE}^{-1}$ terminates as a Taylor series, by only ``looking'' at the the shape of the~HS.

Suppose~$X=\{f_1=0,\ldots, f_c=0\}$ defines a complete intersection in~$\PP_{\CC}^k$ where each~$f_i$ is a homogeneous polynomial of degree~$m_i$ in a standard graded polynomial ring~$R_{k+1}=\CC[X_0,\ldots,X_k]$, such that each variable~$X_i$ has degree~$1$. Then the~HS of~$X$ takes the form
\begin{equation}\label{eq:HS_CI}
\frac{\left(1-t^{m_1}\right)\dots\left(1-t^{m_c}\right)}{(1-t)^n}=\frac{1+a_1t+\dots+a_dt^d}{(1-t)^n}.
\end{equation}
This follows by induction on the~$f_i$ using the additivity of~HS and the the exact sequences
\[
0\to R_k^{[m_i]} \xrightarrow{\cdot f_i} R_k \to R_{k+1} \to 0
\]
where~$R_k^{[m_i]}$ denotes a standard graded polynomial ring with degrees shifted by~$m_i$ so that the first map becomes a morphism of graded rings. Notice~$X$ is a projective variety of codimension~$c$ in~$\PP_{\CC}^k$, i.e.\ has dimension~$\dim = k-c$.

This time~$10\,000$ fake~HS of the form~\eqref{eq:HS_CI} representing~CIs were uniformly drawn from a sample space given by~$c = 1,\ldots,10$,~$m_i = 2,\ldots,10$ and~$1\le n-\sum_i m_i\le11$.
The fake~HS representing non-CI were generated by drawing fake~CI~HS~$f$ from the sample space above and then adding or subtracting a binomial to the numerator preventing the result to factor as in~\eqref{eq:HS_CI}. More precisely, the fake non-CI~HS were computed by
\[
f+(-1)^\varepsilon \cdot \frac{t^{k_0} + (-1)^c\cdot t^{d-k_0}}{(1-t)^n}
\]
where~$\varepsilon=0,1$ and~$k_0=1,\dots,d-1$ was randomly chosen. This procedure ensured that learning is non-trivial, because the resulting fake non-CI~HS have a similar shape to the form~\eqref{eq:HS_CI}, but do not correspond to fake~HS of~CI. The full dataset was comprised by~$10\,000$ fake~CI~HS and~$10\,000$ fake non-CI~HS, i.e.\ a total of~$20\,000$ samples.

We use quotients of successive coefficients in the Taylor expansion of the fake~HS as input to see if the machine could identify complete intersections, i.e.\ we use
\[
\{h_i/h_{i+1} \mid i=0, \dots, n\}
\]
where~$h_i$ denotes the~$i$-th coefficient in the Taylor expansion of the fake~HS and~$n$ is the number of coefficients used. We use~PCA to reduce the dimension followed by a random forest classifier or a~NN. Table~\ref{CIRandomForestClassifier} summarises the averaged accuracies and~MCCs, with standard error, over $10$-fold cross-validations (training performed on the~$10\%$ chunks).

\begin{table}[tb]
\centering
\setlength{\tabcolsep}{0.5em}
\begin{tabular}{cccc}
\toprule
ML & Orders & \multicolumn{2}{c}{Performance Measures}\\
algorithm & of Input & Accuracy & MCC\\
\cmidrule{1-4}
\multirow{2}{*}{PCA+NN} & $0$ to~$100$ & $0.762 \pm 0.010$ & $0.544 \pm 0.030$ \\
& $0$ to~$300$ & $0.951 \pm 0.005$ & $0.902 \pm 0.010$ \\
\cmidrule{1-4}
\multirow{2}{*}{PCA+RF} & $0$ to~$100$ & $0.806 \pm 0.016$ & $0.615 \pm 0.031$ \\
& $0$ to~$300$ & $0.965 \pm 0.003$ & $0.930 \pm 0.005$\\
\bottomrule
\end{tabular}
\caption{Averaged accuracy and~MCC, with standard error, of the~$10$-fold cross-validation of the~PCA+random forest (resp.\ of the~PCA+NN) learning complete intersections in the form~\eqref{eq:HS_CI} with fake~HS coefficients in the specified ranges of degrees as input.}
\label{CIRandomForestClassifier}
\end{table}

If we truncate the Taylor series at order~$100$ and train on~$10\%$ of the data, the accuracy is~${\sim}0.80$ with~MCC~${\sim}0.61$. However, including higher and higher orders of coefficients results into more and more improved results (where the increase in improvement stagnates for sufficiently high orders). For example, if we use Taylor expansions to order~$300$ and train on~$10\%$ of the data, the~PCA+random forest model could give over~$0.95$ accuracy and over~$0.9$~MCC. More precisely, a $10$-fold cross validation (with training performed on the~$10\%$ chunks) would give~$0.965(\pm0.002)$ accuracy (with~$95\%$ confidence interval). We can reproduce these results by using~PCA and a feed-forward~NN with~$4$ hidden dense layers of~$32$ neurons each, dropout layers between the dense layers (dropout factor~$0.05$), LeakyReLU activation (with~$\alpha=0.01$), binary cross-entropy loss function and Adam optimiser. A $10$-fold cross validation with the same input (training performed on the~$10\%$ chunks) yields~$0.951(\pm0.004)$ accuracy (with~$95\%$ confidence interval).

\begin{figure}[!h]
\centering
\includegraphics[width=7cm]{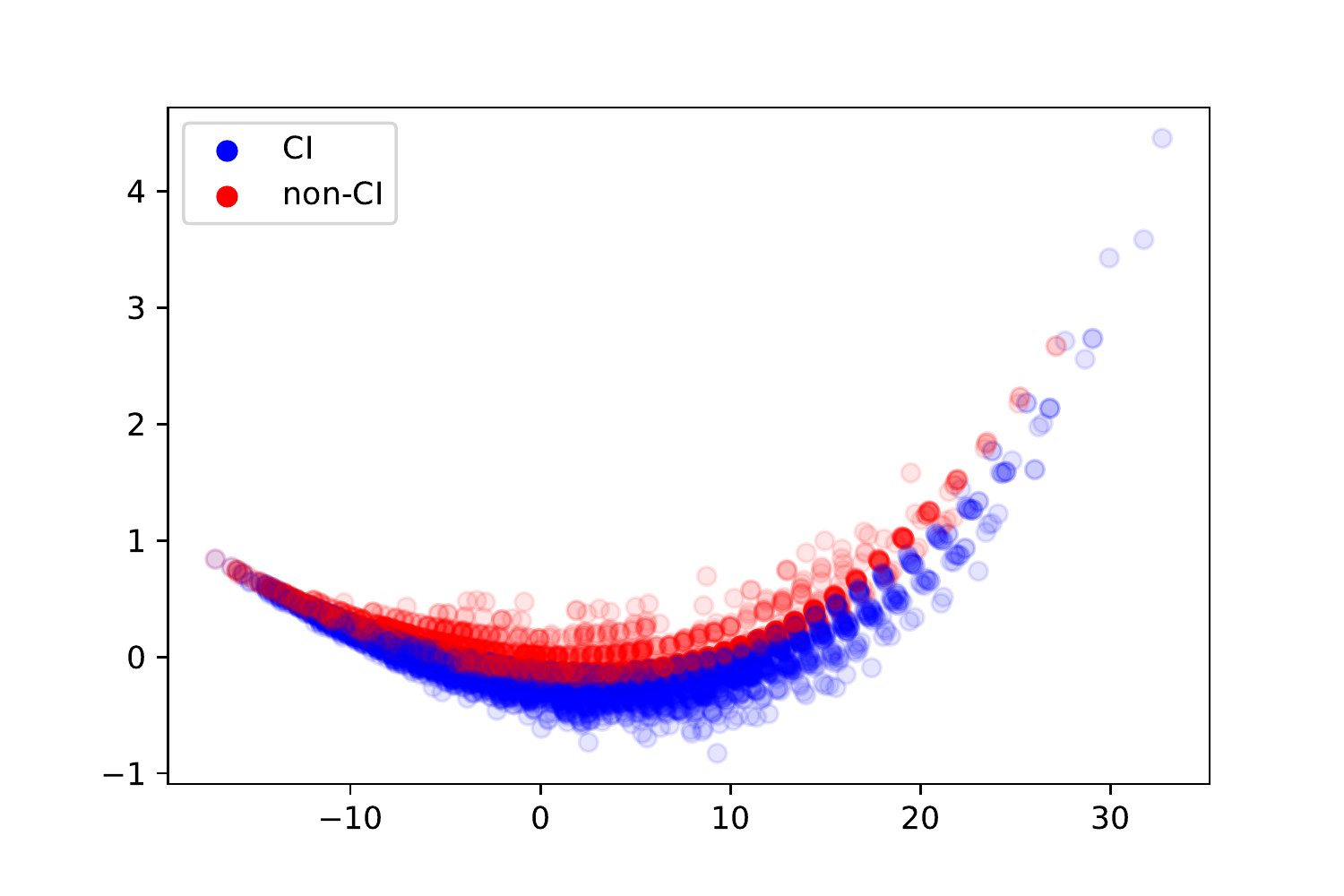}
\caption{PCA for complete and non-complete intersections (with successive quotients of Taylor coefficients).}
\label{ci_pca}
\end{figure}

PCA shows a clear separation of~CIs and non-CIs (see Figure~\ref{ci_pca}). The explained variance ratios show one dominant component with eigenvalue 0.98, where this component has roughly equal contributions from all the series coefficients. This raises the question if this implies that~PCA can efficiently separate~CI from non-CI (\emph{real})~HS or if this is an artefact of our data generation. With 20000 samples of CIs and real non-CIs, we find that a random forest could give $\sim0.8$ accuracy and $\sim0.6$ MCC for a 10-fold cross validation. Although this is a decent result, it would be natural to investigate in future whether there could be better techniques/algorithms to improve such performance. Further study is also necessary to confirm that~PCA is an effective discriminator between~CI and non-CI in this case.

\myparagraph{Acknowledgements.}
JB~is supported by a~CSC scholarship.
YHH~would like to thank~STFC for grant~ST/J00037X/1.
EH~would like to thank~STFC for a~PhD studentship.
JH~is supported by a Nottingham Research Fellowship.
AK~is supported by~EPSRC Fellowship~EP/N022513/1.
SM~is funded by a~SMCSE Doctoral Studentship.
This collaboration was made possible by a Focused Research Workshop grant from the Heilbronn Institute for Mathematical Research.
\appendix
\section{The Plethystic Programme}\label{ap:pleth}
For a function~$f(t)=\sum\limits_{n=0}^\infty a_nt^n$, we can define the~\emph{plethystic exponential} (sometimes known as the Euler transform) as
\[
\text{PE}[f(t)]\coloneqq\exp\left( \sum_{n=1}^\infty \frac{f(t^n) - f(0)}{n} \right) = {\prod\limits_{n=1}^\infty (1-t^n)^{-a_n}}.
\]
For instance, the mesonic~BPS operators fall into two categories: single- and multi-trace. Then the~HS is the generating function for counting the basic single-trace invariants. Moreover, the~HS of the~$N$-th symmetric product is given by~$g_N(t; M) = f(t;{\rm sym}^N(X)),~{\rm sym}^N(X) := M^N/S_N \ ,$ where the ``grand-canonical'' partition function is given by the fugacity-inserted plethystic exponential of the Hilbert series:~$\text{PE}\nu[f(t)] := \prod\limits_{n=0}^{\infty} {(1 - \nu \, t^n)^{-a_n}} = \sum\limits_{N=0}^\infty g_N(t) \nu^N$. In gauge theory, this is considered to be at finite~$N$ and the expansion~$g_N(t) = \sum\limits_{n=0}^\infty b_n t^n$ gives the number~$b_n$ of operators of charge~$n$.

There is also an analytic inverse function to~PE, which is the~\emph{plethystic logarithm}, given by
\[
\text{PE}^{-1}[g(t)] = \sum_{k=1}^\infty
\frac{\mu(k)}{k} \log (g(t^k)), 
\]
where~$\mu(k)$ is the M\"obius function.
The first positive terms in the Taylor expansion of~PE$^{-1}$ encodes generators at different degrees, and the first negative terms give the relations among them. Higher order terms are known as the syzygies. In particular, if~$X$ is a complete intersection, then~$\text{PE}^{-1}[H(t)]$ is a polynomial of~$t$ (i.e. terminates at a finite order).

\vspace{0.01cm}
\section{Real~HS parameter distributions}\label{real_HS_distributions}
\begin{figure}[!htb]
    \centering
    \includegraphics[width=\linewidth]{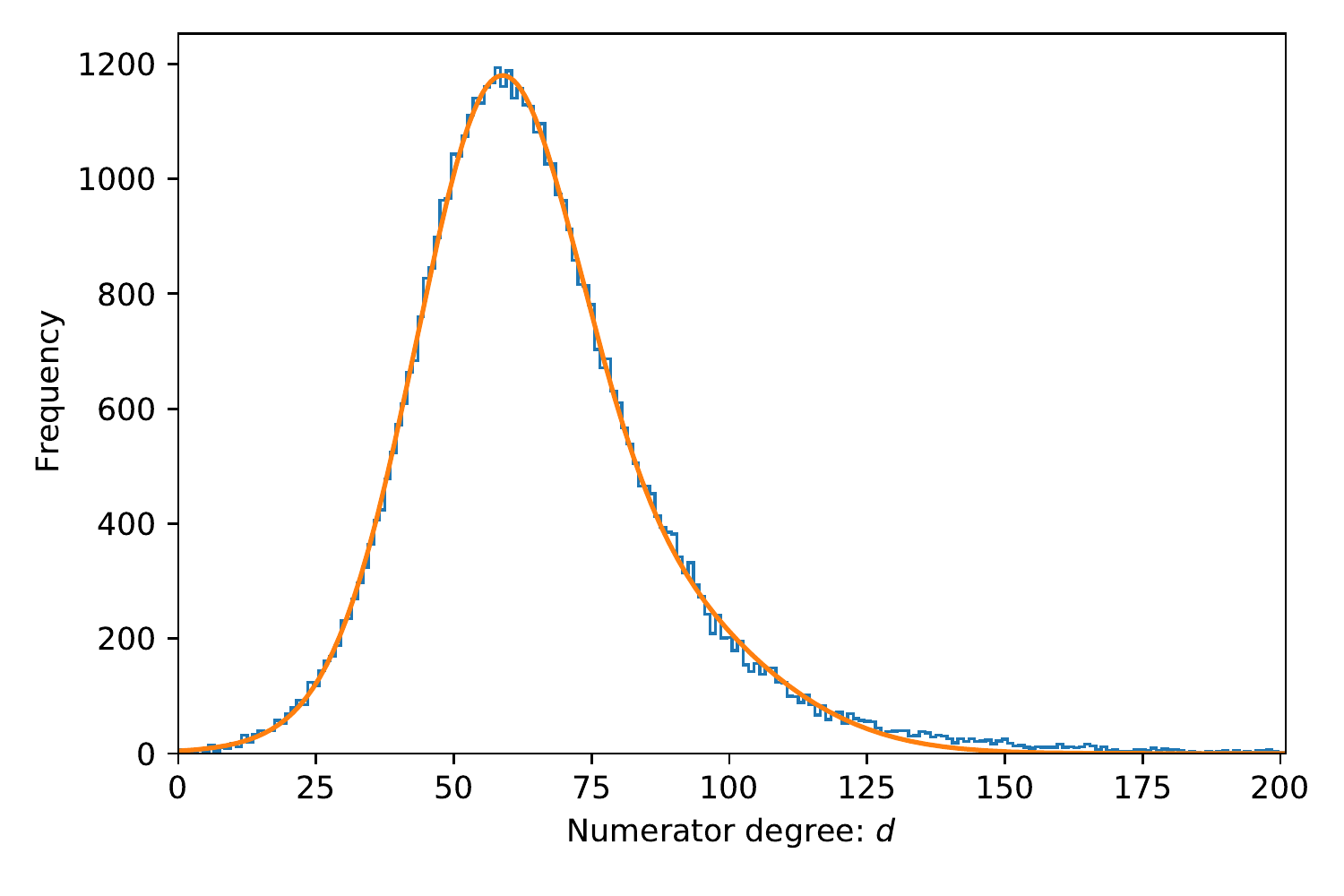} 
    \caption{Histogram of distribution of real~HS numerator degrees~$d$, with Gaussian fitting.}
    \label{num_degrees}
\end{figure}

\begin{figure}[!htb]
    \centering
    \includegraphics[width=\columnwidth]{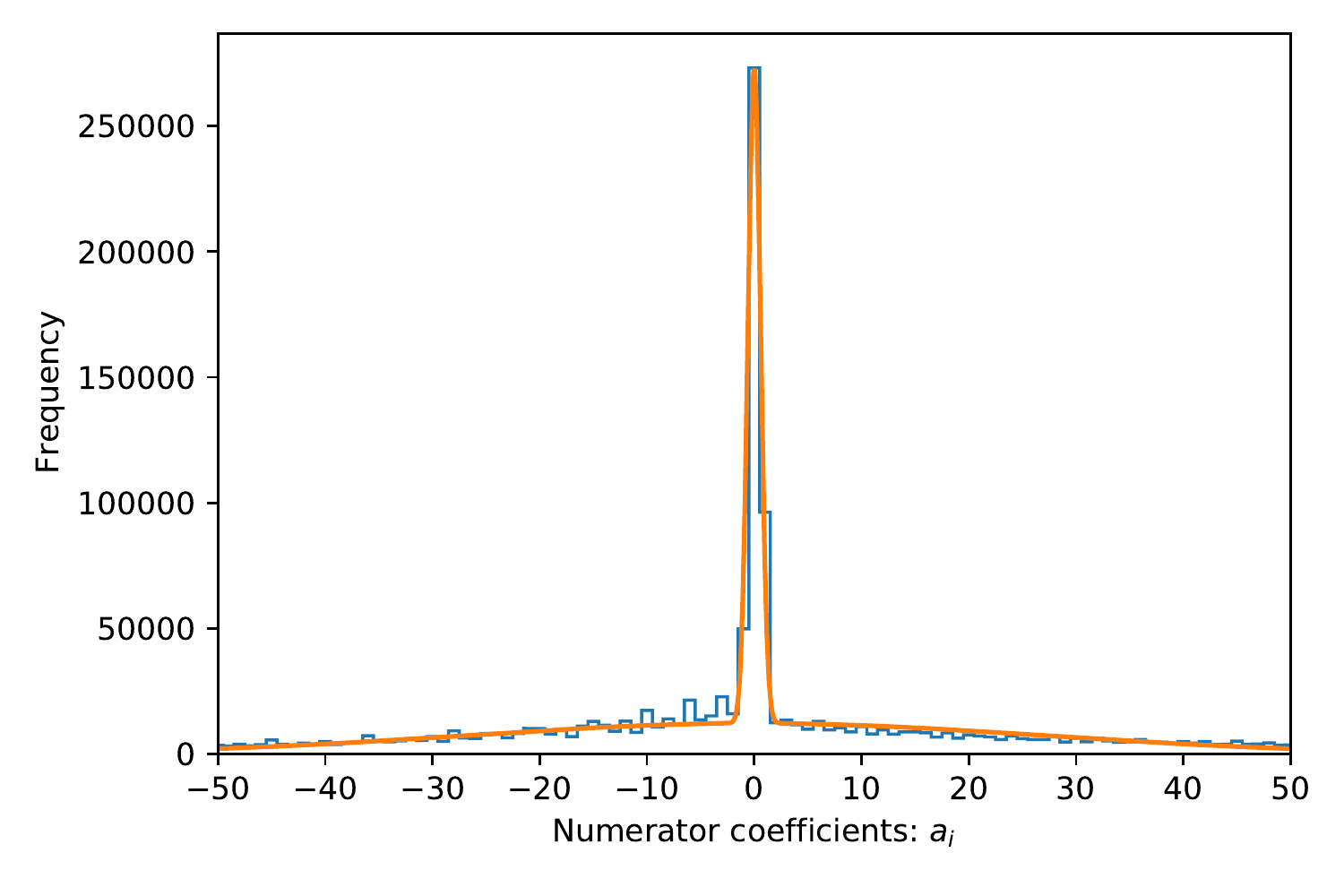} 
    \caption{Histogram of distribution of real~HS numerator coefficient values~$a_i$, with Gaussian fitting.}
    \label{num_coefficients}
\end{figure}

\begin{figure*}[!htb]
\centering
\begin{minipage}[t]{\columnwidth}
\includegraphics[width=\linewidth]{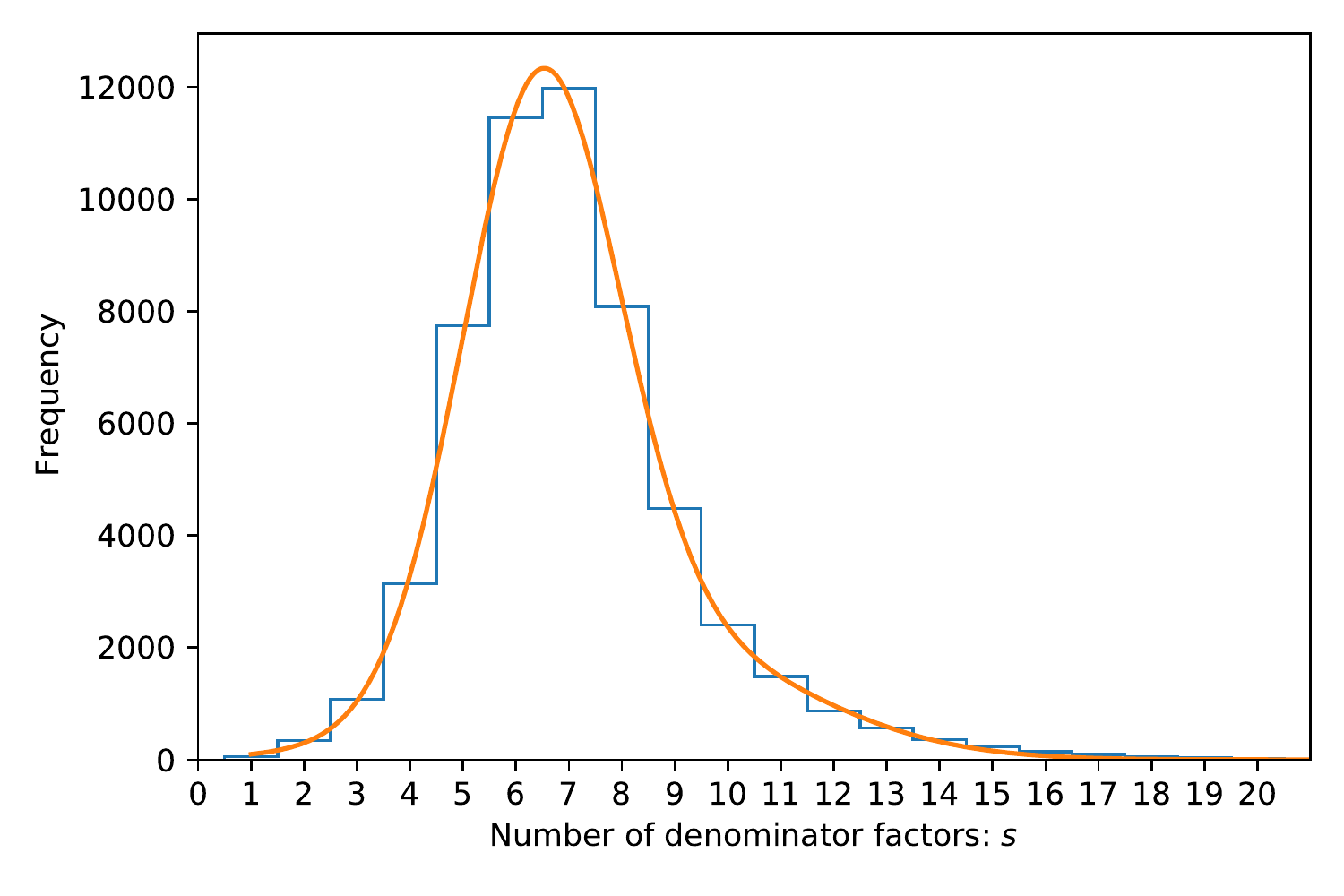} 
\caption{Histogram of distribution of real~HS number of denominator factors~$s$, with Gaussian fitting.}
\label{denom_factors}
\end{minipage}
\hfill
\begin{minipage}[t]{\columnwidth}
\includegraphics[width=\linewidth]{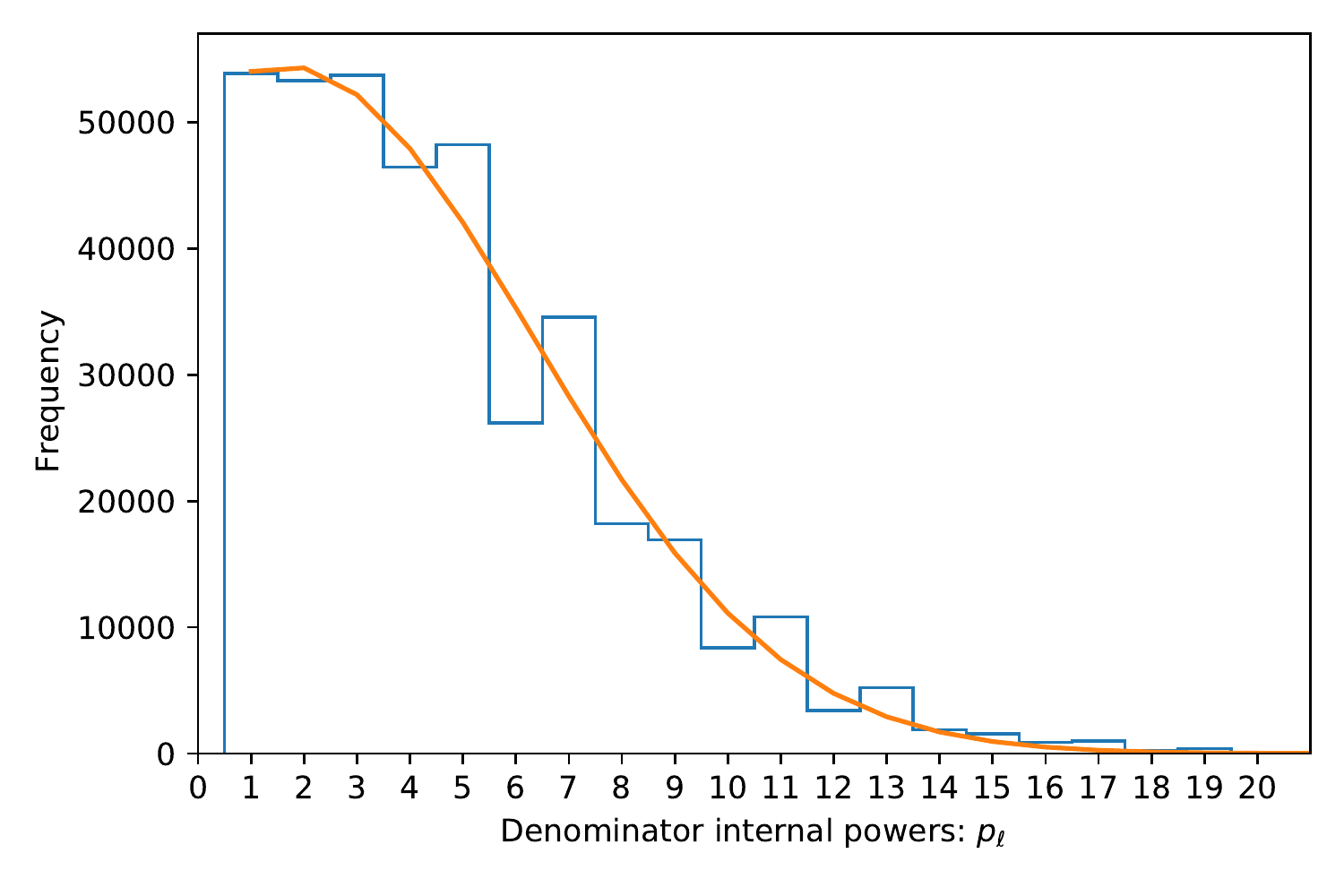} 
\caption{Histogram of distribution of real~HS denominator internal powers (i.e.\ denominator weights)~$p_\ell$, with Gaussian fitting.}
\label{denom_internal_powers}
\end{minipage}
\end{figure*}

\begin{figure}[!htb]
\centering
\includegraphics[width=\columnwidth]{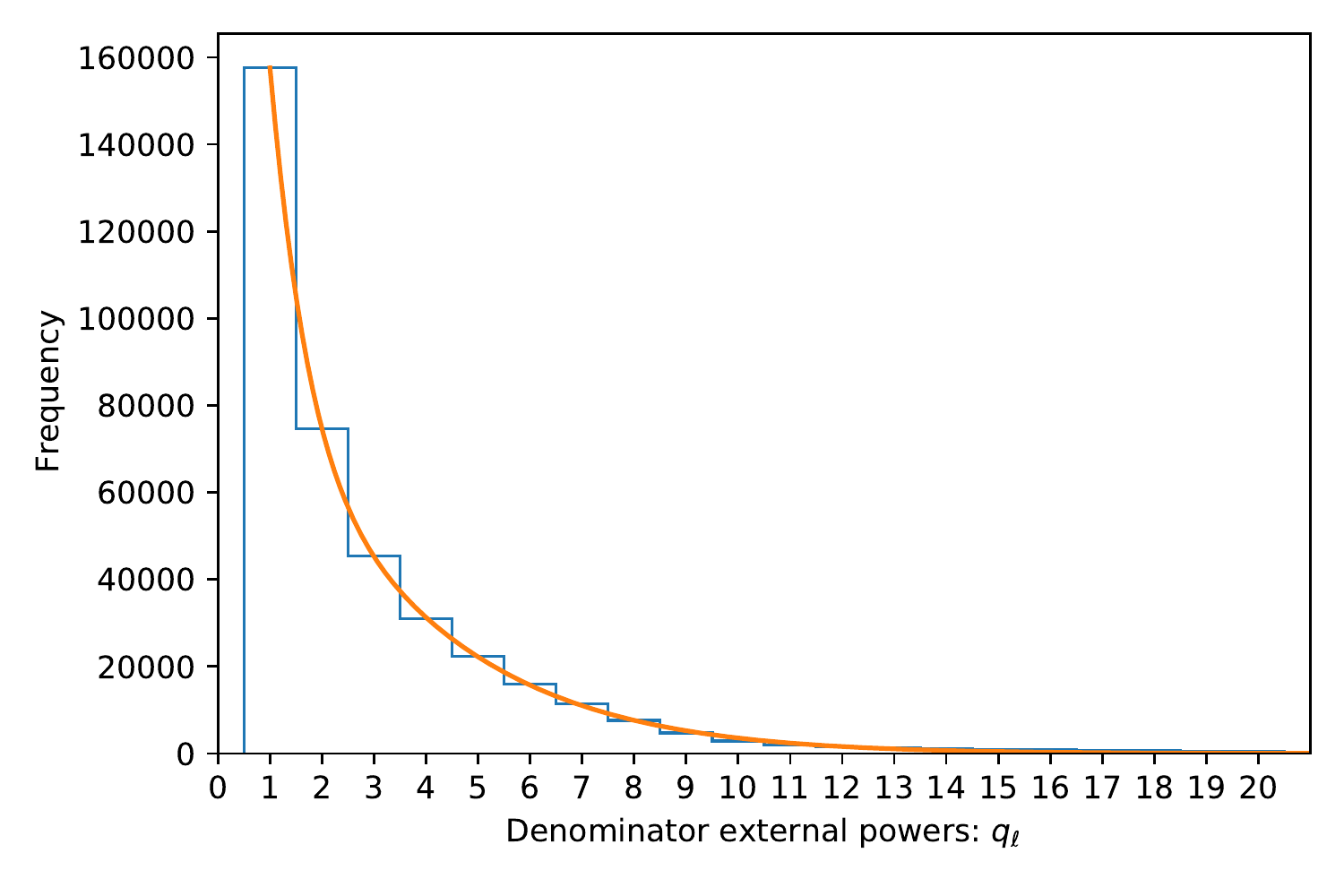} 
\caption{Histogram of distribution of real~HS denominator external powers (i.e.\ number of repetitions of each denominator weight)~$q_\ell$, with Gaussian fitting.}
\label{denom_external_powers}
\end{figure}

\vspace{-0.2cm}
The dataset of real~HS associated to $3$-dimensional Fano varieties considered in this paper~\cite{grdb} that was analysed to produce distributions of the~HS function form parameters~$d, \{a_i\}, s, \{p_\ell\}, \{q_\ell\}$ as shown in Figures~\ref{num_degrees}-\ref{denom_external_powers}.
These distributions, and their respective fittings were used to make fake~HS generation more representative of the real~HS data. 

Fittings used sums of Gaussian distributions, reflecting a Central Limit Theorem motivation in analysis of this large dataset of~${\sim}54\,000$~HS. In all cases the sum of~$2$ independent Gaussian distributions sufficed in making a visually accurate fit. Thus, using these distribution in fake~HS generation would ideally produce~HS of the same form. Interestingly, the fake~HS still had quite different coefficient growth rates to the real~HS, stabilising deeper in the series. This phenomena is further discussed in~\S\ref{detectreal}.


\bibliographystyle{utphys}
\bibliography{bibliography}
\end{document}